\documentclass[twocolumn]{aa}
\usepackage{graphicx}

\newcommand{\gr}{$\gamma$-ray \,}
\newcommand{\grs}{$\gamma$-rays \,}
\begin{document}
   \title{Magnetic Field Amplification in 
          Tycho and other Shell-type Supernova Remnants}

   \author{H. J. V\"olk
           \inst{1}
	   \and
           E. G. Berezhko
           \inst{2}
           \and
           L. T. Ksenofontov
           \inst{2}
           }

   \offprints{H. J. V\"olk}

   \institute{Max-Planck-Institut f\"ur Kernphysik,
                Postfach 103980, D-69029 Heidelberg, Germany\\
              \email{Heinrich.Voelk@mpi-hd.mpg.de}
           \and
              Yu. G. Schafer Institute of Cosmophysical Research and Aeronomy,\\
                     31 Lenin Ave., 678980 Yakutsk, Russia\\
              \email{berezhko@ikfia.ysn.ru}
               \email{ksenofon@ikfia.ysn.ru}             
            }

   \date{Received  / Accepted }

   \abstract{It is shown that amplification of the magnetic field in
supernova remnants (SNRs) occurs in all six objects where morphological
measurements are presently available in the hard X-ray continuum at
several keV.  For the three archetypical objects (\object{SN~1006},
\object{Cas~A} and
\object{Tycho's SNR}) to which nonlinear time-dependent acceleration theory has
been successfully applied up to now, the global theoretical and the local
observational field strengths agree very well, suggesting in addition that
all young SNRs exhibit the amplification effect as a result of very
efficient acceleration of nuclear cosmic rays (CRs) at the outer shock.
Since this appears to be empirically the case, we may reverse the argument
and consider field amplification as a measure of nuclear CR acceleration
and it has indeed been argued that acceleration in the amplified fields
allows the CR spectrum from SNRs to reach the knee in the spectrum or, in
special objects, even beyond. The above results are furthermore used to
investigate the time evolution of field amplification in young SNRs.
Although the uncertainties in the data do not allow precise conclusions
regarding this point, they rather clearly show that the ratio of the
magnetic field energy density and the kinetic energy density of gas flow
into the shock is of the order of a few percent if the shock speed is high
enough $V_\mathrm{s}>10^3$~km/s, and this ratio remains nearly constant during the
SNR evolution. The escape of the highest energy nuclear particles from
their sources becomes progressively important with age, reducing also the
cutoff in the $\pi^0$-decay gamma-ray emission spectrum with time after
the end of the sweep-up phase. Simultaneously the leptonic gamma-ray
channels will gain in relative importance with increasing age of the
sources.

   }
   
  \authorrunning{V\"olk, Berezhko \& Ksenofontov} 
  
  \titlerunning{Field amplification in SNRs}

   \maketitle
%

\section{Introduction}

Recent observations with the Chandra and XMM-Newton X-ray space 
telescopes have confirmed earlier detections of nonthermal continuum 
emission in hard X-rays from young shell-type supernova remnants 
(SNRs). With Chandra it became even possible to resolve spatial scales 
down to the arcsec extension of individual dynamical structures like 
shocks. Results of this type have been published for \object{Tycho's
SNR} (\object{G120.1$+$1.4}) (Hwang  et al. \cite{Hwang}; Bamba et al.
\cite{bam03b}), \object{RCW~86}  (\object{G315.4$-$2.3}) (Rho et al.
\cite{rho02}; Bamba et al. \cite{bam03b}), \object{Cas A} 
(\object{G111.7$-$2.1}) (Vink \& Laming \cite{vl03};  Bamba et al.
\cite{bam03b}), \object{SN 1006}  (\object{G327.6$+$14.6}) (Long et al.
\cite{long}; Bamba et al. \cite{bam03a}; \cite{bam03b}; \cite{bam04}),
and \object{Kepler's SNR} (\object{G4.5$+$6.8}) (Bamba et al.
\cite{bam03b}); (see also Bamba \cite{ba04}). Small-scale  filamentary
structures in \object{SNR RX~J1713.7$-$3946}  (\object{G347.3$-$0.5})
have been reported by  Uchiyama et al. (\cite{uch03}) and Lazendic et
al. (\cite{laze}).

These results can be seen from three different viewpoints: First of all,
assuming the filamentary structures to represent synchrotron emission near
shocks, their width can be considered as the synchrotron coBoling length to
derive the effective downstream magnetic field strength $B_\mathrm{d}$.
Secondly, we can compare this $B_\mathrm{d}$ with the value for the interior
magnetic field strength that makes a complete dynamical model of the SNR
consistent with the spatially integrated synchrotron spectrum, from the
radio range to hard X-rays. The third viewpoint is from the plasma physics 
of field amplification in an ionized gas with a very strong nuclear cosmic
ray (CR) gradient.

Let us first turn to the {\it observed X-ray morphology}. In their work on
Cas A, Vink \& Laming (\cite{vl03}) have interpreted some of the filamentary hard
X-ray structures as the result of strong synchrotron losses of the
emitting multi-TeV electrons in amplified magnetic fields downstream of
the outer accelerating SNR shock. Such an interpretation has been
independently given by Berezhko et al. (\cite{bkv02}) from a consistent
description of the observed SN~1006 dynamics and of the properties of its
overall nonthermal emission, and by Berezhko et al. (\cite{bkv03}) for the
Chandra observations of SN 1006, leading to an effective downstream field
strength of 120~$\mu$G for the latter object. Subsequently Berezhko \&
V\"olk (\cite{bv04a}) revised the Vink \& Laming result for Cas A by consistently
taking account of all relevant dynamical and geometrical (projection)
factors. It was shown that for the highest energy electrons, usually
responsible for the X-ray emission (like in the case of SN~1006 and
Cas~A), synchrotron losses become important already during the
acceleration process. Together with the projection effect they imply a
significantly shorter electron cooling time behind the shock, and thus an
effective downstream magnetic field strength in Cas A that is stronger by
a factor of five than estimated by Vink \& Laming (\cite{vl03}). It reaches about
0.5 mG in the immediate downstream region of the outer shock.

Such observational results gain their qualitative significance through the
fact that for SN 1006 (Berezhko et al. \cite{bkv02}) and Cas A (Berezhko et al.
\cite{bpv03}) these effective magnetic fields and morphologies turned out to be
exactly the same as predicted theoretically from {\it nonlinear
acceleration theory}. The theory uses gas dynamics and nonlinear
acceleration theory to calculate a kinetic model of young, evolving SNRs
which includes the energetic electron component together with the nuclear
CR component. Since this theory does not explicitly contain equations for
the evolution of the electromagnetic fields or for the injection of
low-energy particles into the acceleration process, these processes are
parametrized by an effective magnetic field strength $B_\mathrm{d}$ and an
injection strength for the nuclear particles. The theory then works as
follows: the comparison of the model with the observed spectral features
of the spatially integrated synchrotron emission from the remnant yields a
selfconsistent value for $B_\mathrm{d}$ (see also Reynolds \& Ellison \cite{re92}) as well
as for the total CR pressure $P_\mathrm{c}$. The amplitude of the observed
synchrotron emission limits in addition the pressure of the energetic
electrons to 1 percent or less of the total CR pressure $P_\mathrm{c}$ in the
remnant.

It is important to note that the magnetic field $B_\mathrm{d}'$ derived from the
fit of the calculated overall synchrotron flux to the observed flux
and the field $B_\mathrm{d}''$ that results from the observed X-ray brightness
profile correspond to basically different parts of the SNR. The field
$B_\mathrm{d}''$ belongs to the thin region just behind the shock whereas $B_\mathrm{d}'$,
which mainly comes from the radio data, characterizes the average
effective field within a much wider region, occupied by the GeV-electrons.
Generally speaking, $B_\mathrm{d}'$ and $B_\mathrm{d}''$ can therefore also be basically
different. The equality $B_\mathrm{d}'=B_\mathrm{d}''$ is expected if during the SNR
evolution the random amplified field is generated near the shock front at
the level $B_\mathrm{d}''^2 \propto V_\mathrm{s}^2$ (Berezhko et al. \cite{bkv03}; Berezhko \&
V\"olk \cite{bv04b}). Therefore the equality of the experimental values $B_\mathrm{d}'$
and $B_\mathrm{d}''$ confirms that the field is indeed amplified (generated) to the
above level.

It should perhaps be noted here that there is a good semi-quantitative
understanding of the injection rate of nuclear particles into the
diffusive shock acceleration process in terms of the escape of
suprathermal downstream particles. However, due to the incompletely known
relaxation state of the downstream ions, the escape rate is theoretically
known only up to factors of order unity (V\"olk et al. \cite{vbk03}), and
therefore it is necessary to fix its value empirically in a quantitative
model for the SNR as a whole. Overall, with the three parameters $B_\mathrm{d}$,
$P_\mathrm{c}$, and electron-to-proton ratio determined empirically, it is possible
to describe the overall observed electron synchrotron spectrum from first
principles.

The large total CR pressure and the negligible contribution to it from
electrons is a compelling argument for the existence and dominance of the
nuclear CR component in SNRs which makes up almost the entire pressure
$P_\mathrm{c}$. The implied magnetic field strength must be the same as that
determined independently from the observed X-ray filaments. This is a
direct empirical test for acceleration theory. Even though such a detailed
agreement has been demonstrated only in two cases up to now, it is hard to
believe that this is a mere coincidence.

Besides morphology and acceleration theory the third aspect of field
amplification concerns {\it plasma instability theory}. In shocks with
strong nuclear CR production the downstream CR pressure $P_\mathrm{c}$
is of the order of the ram pressure $\rho V_\mathrm{s}^2$ of the upstream gas flow
in the shock frame (Drury \& V\"olk \cite{druV};  Axford et al. \cite{alm}). Then the
unstable streaming of the shock accelerated nuclear particles (Bell
\cite{bell78};
Blandford \& Ostriker \cite{blano}) leads to an excessive resonant excitation of
Alfv\'{e}n waves whose energy density exceeds that of the mean magnetic
field by far (McKenzie \& V\"olk \cite{mcv82}), making an increased and maximally
disordered magnetic field plausible (V\"olk \cite{voelk84}). Much more recently,
analytical studies and numerical simulations of this instability were
performed by Lucek \& Bell (\cite{lucb00}) and Bell \& Lucek
(\cite{bluc01}) which indicated
strong field amplification for $P_\mathrm{c} \sim \rho V_\mathrm{s}^2$, while the spatial
diffusion for energetic particles in this disordered field reached the
limiting level of Bohm diffusion. This implies that the scattering mean
free path approaches the gyro radius in the amplified field. Finally, Bell
(\cite{bell04}) extended an early investigation by Achterberg
(\cite{ach83}) of the
nonresonant right-hand polarized low frequency MHD mode, propagating
parallel to the average magnetic field in the presence of a nuclear CR
current. In the case $P_\mathrm{c} \sim \rho V_\mathrm{s}^2$ he found strong linear growth
with a growth rate in excess of the oscillation frequency, i.e. a purely
growing mode. This growth is even more rapid than that of the resonant
Alfv\'{e}n waves, usually considered until now. Nonlinear quasi-MHD
simulations on the basis of this instability showed strong field
amplification for the energetic particle streaming that one has to expect
in the strong accelerating shocks of young shell-type SNRs.

We assume the observed filamentary X-ray structures to be manifestations
of such theoretical plasma processes and shall therefore consider them as
an integral aspect of diffusive shock acceleration in violent events such
as supernova explosions.

The aim of our paper is to discuss the objects where field amplification
possibly occurs. Therefore we shall consider all those shell-type SNRs
where filamentary shock structures in hard X-rays have been published up
to now. These SNRs are typically quite young, with ages not much in excess
of a sweep-up time for a uniform circumstellar medium. The explanation for
this selection effect is as follows: the cutoff frequency of synchrotron
spectrum produced by the shock accelerated electrons is proportional to
$V_\mathrm{s}^2$ and diminishes after sweep-up as a result of the slowing down of
the shock, reducing the X-ray emission at a fixed observation energy
energy, like 4 keV, sharply with time (Berezhko \& V\"olk \cite{bv04b}).
This is independent of the fact that the total CR energy, the normalized
CR pressure $P_\mathrm{c}/(\rho V_\mathrm{s}^2)$, and the hadronic gamma-ray luminosity
remain quite high for much longer times in the Sedov phase (e.g. Berezhko
\& V\"olk \cite{bv97}). Older SNRs are therefore not expected to exhibit
significant nonthermal hard X-ray emission.

To search for field amplification in young SNRs only does therefore not
imply a methodological restriction. On the contrary, if we knew that all
young SNRs showed field amplification we would know that all SNRs have a
strong nuclear energetic particle population, as a result of the strong
theoretical connection between field amplification and the dominant
contribution of nuclear CRs to the overall nonthermal pressure $P_\mathrm{c}$.

Besides \object{Cas A} and \object{SN 1006} the available Chandra observations concern
\object{Tycho's} and \object{Kepler's SNR}s, \object{RCW~86}, and
\object{RX~J1713.7-3946}, as mentioned
above. \object{Tycho's SNR} plays a special role, because there exists a detailed
theoretical model for the particle spectra and the synchrotron morphology
(V\"olk et al. \cite{vbkr02}). We shall use the
 projection model for the
brightness morphology developed by Berezhko \& V\"olk (\cite{bv04a}) wherever
possible to fit the data by adjusting the strength of the effective
magnetic field $B_\mathrm{d}$. We find that all the investigated sources have
amplified fields to a varying degree, and therefore we propose that all
these sources are efficient nuclear CR accelerators. This is a necessary
condition for the Galactic SNRs to constitute the {\it source population}
of the Galactic CRs, at least up to the so-called knee of the energy
spectrum. Regarding \object{Tycho's SNR} we shall find that it constitutes the
third case where spectral and morphological synchrotron characteristics
independently give the same effective field strength.

\section{Effective magnetic fields from X-ray data}

As shown by Berezhko \& V\"olk (\cite{bv04a}), the width $L\approx 7 l_2$ of the
observable brightness profile $J_{\nu}(\epsilon_{\nu}, \rho)$ is always
appreciably wider than the downstream width $l_2$ in radius $r$ of the
three-dimensional emissivity $q_2 (\epsilon_{\nu},r)$, simply for 
geometric reasons:
\[
J_{\nu}=2q_2\frac{R_\mathrm{s}l_2}{\sqrt{R_\mathrm{s}^2-l_2^2}}
\left\{
1-\frac{l_2}{R_\mathrm{s}}
\left(\frac{R_\mathrm{s}^2-2\rho^2}{R_\mathrm{s}^2-\rho^2}\right)-\right.
\]
\[
\hspace{1cm}
-\exp\left(\frac{\rho^2-R_\mathrm{s}^2}{R_\mathrm{s}l_2}\right)
\left[
1-\frac{l_2}{R_\mathrm{s}}
\left(\frac{R_\mathrm{s}^2-2\rho^2}{R_\mathrm{s}^2-\rho^2}\right)+ \right.
\]
\begin{equation}
\left. \left.\hspace{1cm}
+\left(\frac{R_\mathrm{s}^2-\rho^2}{2R_\mathrm{s}l_2}-1\right)
\left(\frac{2\rho^2-R_\mathrm{s}^2}{R_\mathrm{s}^2}\right)
\right]
\right\}.
\label{eq1}
\end{equation}
In this approximate relation, valid for $l_2 \ll 0.1 R_\mathrm{s}$, $R_\mathrm{s}$ and
$\rho$ are the shock radius and the distance between the center of the
remnant and the line of sight, respectively. At the same time $l_2$ is the
distance through which the very energetic electrons stream away from the
shock during a synchrotron loss time $\tau(p)= 9 m_\mathrm{e}^2c^2/(4
r_0^2B_\mathrm{d}^2p)$, where $m_\mathrm{e}$ is the electron mass, $r_0\approx 2.8 \times
10^{-13}$cm is the classical electron radius, $c$ denotes the speed of
light, and $B_\mathrm{d}$ is the effective magnetic field in the downstream region.
For given frequency $\nu$ the momentum $p$ of the radiating electron is
approximately given as $p\propto \sqrt{\nu/B_\mathrm{d}}$.

In the limit of a strongly fluctuating magnetic field around the shock on
all scales, we also assume that the diffusion coefficient $\kappa(p)$ is
given by the Bohm limit, $\kappa(p)=\rho_\mathrm{B}v/3$, where
$\rho_\mathrm{B}$ and $v$ are
the particle gyroradius and velocity, respectively. Therefore $l_2$ is
determined by $B_\mathrm{d}$ and $\nu$ which can be conveniently turned into an
equation for $B_\mathrm{d}$ in terms of $l_2$ and $\nu$ (Berezhko \& V\"olk 
\cite{bv04a}):
\begin{equation}
B_\mathrm{d}=[3m_\mathrm{e}^2c^4/(4er_0^2l_2^2)]^{1/3}(\sqrt{1+\delta^2} -\delta)^{-2/3},
\label{eq2}
\end{equation}   
with
\begin{equation}
\delta^2=0.12[c/(r_0\nu)][V_\mathrm{s}/(\sigma c)]^2,
\label{eq3}
\end{equation}  
where $e$ is the proton charge. 
These are the main relations which we
are going to use in order to determine the amplified field. The 
Chandra observations yield $J_{\nu}(\epsilon_{\nu}, \rho)$ in terms of
the angular distance $\Delta \psi = (\rho-R_\mathrm{s})/d$
which can be converted to $\rho$ for
known source distance $d$. Given the $J_\nu (\rho)$-profiles 
for a given X-ray energy $\epsilon_{\nu}$ we shall then fit the
data to the model curve Eq.(\ref{eq1}) to determine $l_2$, and thus
$B_\mathrm{d}$. In some of the published observations, mentioned in the
Introduction, only the exponential widths of the profiles are quoted. In
those cases we shall approximately equate $L$ with this width.

\subsection{\object{Tycho's SNR}} 

In the case of \object{Tycho's SNR} the circumstellar gas density is unexpectedly
inhomogeneous for a type Ia SN. This results in significant azimuthal
deviations from spherical symmetry both in shape and expansion velocity of
the SNR shock (e.g. Reynoso et al. \cite{rvdg99}), even though to first
approximation the remnant can be considered as spherical. There are also
indications that the local radio spectral index $\alpha$, derived from the
ratio of the emissions at $\lambda = 20$~cm and $\lambda = 90$, is
nonuniform (Katz-Stone et al. \cite{Katz00}) and in several locations quite low,
near $\alpha=0.5$.  The well-defined spatially integrated radio spectrum
on the other hand is rather steep ($\alpha \approx 0.61$) if fitted by a
power law, and indicates curvature (Reynolds \& Ellison \cite{re92}). It is
therefore perhaps not surprising that also the radial brightness profiles
across the shock are not identical in different regions of the shock
circumference. Two such profiles were presented by Hwang et al. (\cite{Hwang}).

There are additional reasons for different widths of such profiles apart
from inhomogeneities of the circumstellar density. Being an example of a
type Ia SN, the low-mass progenitor star should not have had a significant
mass loss in the form of a wind with a Parker spiral-type circumstellar
field. To the extent that the density of the environment is uniform, also
the ambient magnetic field should be uniform, ideally resulting in a
dipolar topology of the distribution of accelerated nuclear particles
(V\"olk \cite{vo97}; V\"olk et al. \cite{vbk03}). This should also lead to a
corresponding distribution of field amplification in azimuth and thus to
different spatial scales of the radial synchrotron profiles. Even though
such a simple geometry is not realized in \object{Tycho's SNR}, the
underlying physical processes should nevertheless lead to azimuthal
variations of the radial synchrotron emissivity scales in a
quasi-statistical way.

As indicated in the Introduction a global value of the effective interior
magnetic field -- and thus the sharpness of the radial profile at the
shock -- can be independently determined from a comparison of the
volume-integrated theoretical synchrotron spectrum with the one observed.  
In fact, the nonlinearity of the diffusive acceleration process for CR
nucleons leads to a spectral steepening and even to a curvature of the
energy spectrum of accelerated electrons for rigidities corresponding to
sub/trans-relativistic protons. This well-known feature (e.g. Ellison \&
Eichler \cite{elei}; Drury \cite{drury83}; V\"olk \cite{voelk84};
Blandford \& Eichler \cite{blanei}; Reynolds
\& Ellison \cite{re92}; Berezhko et al. \cite{byk96}) implies a steeper and curved
synchrotron spectrum in the radio range compared to the test particle
limit, if the radiating electrons are indeed in the above low rigidity
range. An observed integrated spectrum with this feature then implies a
minimum magnetic field strength. But this is only the most obvious
nonlinear feature. In fact, the entire frequency range of the synchrotron
spectrum must be consistent with this magnetic field strength, especially
also in the hard X-ray region. This fixes both the amount of shock
modification, i.e. the proton injection rate in the theory, and the global
effective magnetic field strength (Berezhko et al. \cite{bkv02}). It is remarkable
that with these three parameters it is possible to describe the
measurements of the entire synchrotron spectrum, which in the case of 
Cas~A include also the mm and mid-infrared range.

As a consequence we have two experimentally independent methods to
determine the effective magnetic field, a local one and a global one.
Several independent radial profiles even yield several independent local
values as a function of azimuth. We shall in the following investigate the
consistency of these values for \object{Tycho's SNR}.

The radial width $l_2$ of the three-dimensional emissivity distribution is
smaller by a factor of about 7 than the radial brightness scale of the
optically thin emission. It is simple to verify that for the considered
X-ray energies, which lie in the cutoff region of the integrated
synchrotron brightness, the case of strong losses (V\"olk et al.
\cite{vmf81}) is
relevant, where the emissivity width is given by the limiting case of
Eqs.(\ref{eq2}, \ref{eq3}) (Berezhko \& V\"olk \cite{bv04a}) 
\begin{equation} 
l_2=\sqrt{\kappa_2 \tau_2}, 
\label{eq4}
\end{equation} 
i.e. by the postshock {\it diffusion length} in a
synchrotron loss time.  In the Bohm diffusion limit for the accelerating
particles in \object{Tycho's SNR} (which at the present epoch is in the transition
between the free expansion phase and the Sedov phase), $l_2$ is then given
by 
\begin{equation} 
l_2=[3m_\mathrm{e}^2c^4/(4er_0^2B_\mathrm{d}^3)]^{1/2}. 
\label{eq5}
\end{equation} 

In other words, the radial emissivity profile is
parametrized by $B_\mathrm{d}$ alone in this limit.

We shall start with the Chandra data that were given in Fig.~4 of Hwang et
al. (\cite{Hwang}) in the form of radial profiles of the 4-6 keV continuum
brightness for two azimuthal sectors in the northwest (NW) and the
southwest (SW). This energy range excludes most of the emission lines. The
profiles show a pronounced outer rim, presumably at the location of the
remnant's forward shock.

In the lowest approximation the narrower rim in the NW has an exponential
width of about 3.5 arcsec, whereas the SW rim extends over at least 4.7
arcsec. This corresponds to about 1.6 and 2.1 percent of projected radius.
Adopting a distance of $d=2.3$~kpc and using $l_2\approx L/7$, this
implies $l_2 \approx (1.7 \pm 0.7) \times 10^{16}$~cm and 
$l_2 \approx (2.3 \pm 0.7) \times 10^{16}$~cm 
for the NW and SW profile, respectively.
Using the full Eqs. (\ref{eq2}) and (\ref{eq3}) with $V_\mathrm{s}=3100$~km/s, and a
shock compression ratio $\sigma = 6$, from V\"olk et al. (\cite{vbkr02}), and
taking $\epsilon_{\nu}=h \nu \approx 4 $~keV, we obtain 
$B_\mathrm{d} \approx 404 \, (+169 -82)$~$\mu$G and $B_\mathrm{d} \approx 332 \, (+ 88 -
53)$~$\mu$G, 
for the two respective sectors. The shock sectors are
obviously independent except possibly the innermost regions near the
center which play no role in the present argument. Nevertheless one can
conclude that the magnetic field values within these regions are roughly
the same within the errors.

Bamba (\cite{ba04}) has given an exponential width  
$L=0.073(+0.010-0.009)$~pc for a profile in the NW, in the 2 - 10 keV 
band. Using $\epsilon_{\nu} = 2 $~keV and $l_2=L/7$ we obtain 
from Eqs. (\ref{eq2}) and (\ref{eq3}) $B_\mathrm{d} =301 (+28 - 25)$~$\mu$G.

\begin{figure} 
\centering 
\includegraphics[width=7.5cm]{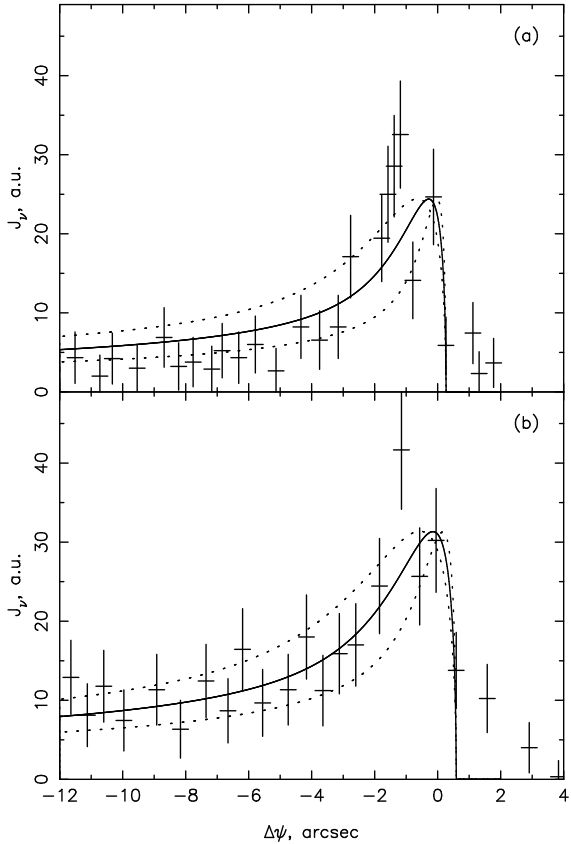}
\caption{The 
Chandra $4-6$~keV continuum brightness radial profile of \object{Tycho's
SNR}
observed by Hwang et al. (\cite{Hwang}).  
Upper panel (a): southwest azimuthal sector. Lower panel (b): northwest
azimuthal sector. The statistical error in the brightness values was assumed to
equal the square root of the values, whereas for the angular position an
error of $0.5''$ was assumed. The
data are fitted to the projection of the exponential emissivity profile
described by the model of Eq. (\ref{eq1}) in the downstream region {\it (solid
line)}. The fit has a $\chi^2 = 17.3$, with $\chi^2/dof = 1.73$
and $\chi^2 = 6.56$, with $\chi^2/dof = 0.82$ for (a) and (b) respectively. 
The {\it
dotted lines} indicate the $1\sigma$ deviations.}
\label{f1}
\end{figure}

We have also fitted both the widest and most narrow template
distributions, cf. Eqs. (\ref{eq1}--\ref{eq3}), to the Hwang et al. profiles. To do this we
assumed a Gaussian distribution of the statistical errors in the
measurement of the brightness distribution $J_{\nu}$ , i.e. the errors to
be given by the square root of the discrete values given in the paper by
Hwang et al. (\cite{Hwang}), together with a positional accuracy of the ACIS array
of Chandra of 0.5 arcsec. The results for the two radial profiles are
given in Fig.~\ref{f1}. This implies $B_\mathrm{d}= 331(+182-94)$~$\mu$G for the SW
profile, and $B_\mathrm{d}=272(+117-70)$~$\mu$G for the profile in the NW. Note
that the values $\chi^2/dof = 1.73$ and $\chi^2/dof = 0.82$ for the cases
represented in Fig.~\ref{f1}a and \ref{f1}b respectively show that the data are fitted by
the formula (\ref{eq1}) quite well. Again one can see that within the
uncertainties the field values in these two region are the same.

\begin{figure} 
\centering 
\includegraphics[width=7.5cm]{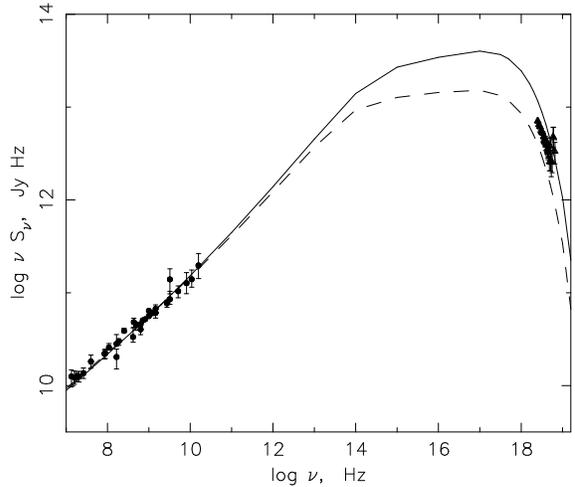}
\caption{Synchrotron spectral energy density $\nu S_{\nu}$ of
\object{Tycho's SNR}
as a function of frequency $\nu$. The {\it solid curve} corresponds to a
fit with a downstream magnetic field of $B_\mathrm{d}=240$~$\mu$G (V\"olk et al. \cite{vbkr02})
which slightly overestimates the X-ray continuum at $\nu > 10^{18}$~Hz and
therefore underestimates the field. The more strongly nonlinearly modified
case, with slightly steeper radio spectrum ({\it dashed curve}) requires 
a
higher B-field of $B_\mathrm{d}=360$~$\mu$G but somewhat underestimates the X-ray
brightness. The actual B-field value lies in between these cases.}
\label{f2}
\end{figure}

For the global determination of the effective downstream field $B_\mathrm{d}$ we
compare the theoretical synchrotron spectrum with the observed spatially
integrated spectrum (see Fig.~\ref{f2}). To explore the possible bandwidth of $B_\mathrm{d}$-values we
approximate the data with a slightly softer radio spectrum than adopted by
V\"olk et al. (\cite{vbkr02}). In the framework of the strongly nonlinear system of
equations, the required stronger shock modification by the nuclear
component is achieved by a somewhat increased proton injection rate. The
theoretically implied increase of the magnetic field from 240 to 360
$\mu$G then requires a reduction of the amplitude of the electron momentum
distribution since the radio electrons are not subject to synchrotron
cooling at the present epoch. This leads to a decreased electron/proton
ratio in the energy range where radiative cooling is unimportant. Most
importantly however, the cooling region of the synchrotron emission
spectrum $\nu>10^{14}$~Hz will be lowered in amplitude by this decreasing
electron/proton ratio.

Taking into account, that the electron distribution function is roughly a
power law $f_\mathrm{e}\propto p^{-q}$ we can approximately write
\begin{equation}
S_{\nu}\propto p^3 f_\mathrm{e}(p)B_\mathrm{d} l_2
\end{equation}
for $\nu \approx (p/mc)^2 eB_\mathrm{d}/mc$, where $f_\mathrm{e}(p)$ is the
spatially averaged electron distribution in the remnant interior. The low
frequency part of the spectrum $S_{\nu}(\nu<10^{14}$~Hz) is produced by
electrons which do not suffer synchrotron losses. Therefore $l_2\sim 0.1
R_\mathrm{s}$ and $S_{\nu}\propto \eta K_\mathrm{ep}\nu^{-\alpha}B_\mathrm{d}^{\alpha+1}$, where
$\eta$ is proton injection rate, $K_\mathrm{ep}$ is the electron-to-proton ratio,
and $\alpha=(q-3)/2$. To have an equally good fit for the radio data the
electron-to-proton ratio $K_\mathrm{ep}$ should be reduced for a higher
downstream field value $B_\mathrm{d}$.

At higher frequencies $\nu>10^{14}$~Hz the synchrotron emission is
produced by electrons which suffer significant synchrotron losses. For
$\nu<\nu_\mathrm{max}=10^{16}$~Hz $l_2\approx u_2\tau_2$, which corresponds to the
so-called weak loss case (Berezhko \& V\"olk \cite{bv04a}). Since $\tau_2\propto
p^{-1}B_\mathrm{d}^{-2}$ we have $S_{\nu}\propto \eta K_\mathrm{ep}p^{2-q}/B_\mathrm{d}\propto \eta
K_\mathrm{ep}\nu^{-\alpha}B_\mathrm{d}^{\alpha-1}$, where $\alpha=(q-2)/2$. Since the
power law index $q$ is close to 4, $S_{\nu}\propto \eta K_\mathrm{ep} \nu^{-1}$,
independently of $B_\mathrm{d}$. The reduction of the electron- to-proton ratio
$K_\mathrm{ep}\propto B_\mathrm{d}^{(q-1)/2}$ for higher field $B_\mathrm{d}$ leads to a decrease
of the overall synchrotron flux $S_{\nu}$ (see Fig.~\ref{f2}).

The reduction of electron
production, discussed above, leads to a proportional reduction of the
synchrotron emission in the cooling range as shown by the dashed curve in
Fig.~\ref{f2} and implies an underestimate of the X-ray synchrotron
emission compared to the slight overestimate in V\"olk et al.
(\cite{vbkr02}, see
Figs.~3 and 5). The spectral fit in that earlier paper had on purpose not
been fine-tuned, because the aim had been to demonstrate field
amplification in principle. The optimum magnetic field strength then lies
between the two cases, roughly at 300~$\mu$G. This leads to an optimum 
value $B_\mathrm{d} \approx 300 \pm 60$~$\mu$G from acceleration theory.

\begin{figure}
\centering
\includegraphics[width=7.5cm]{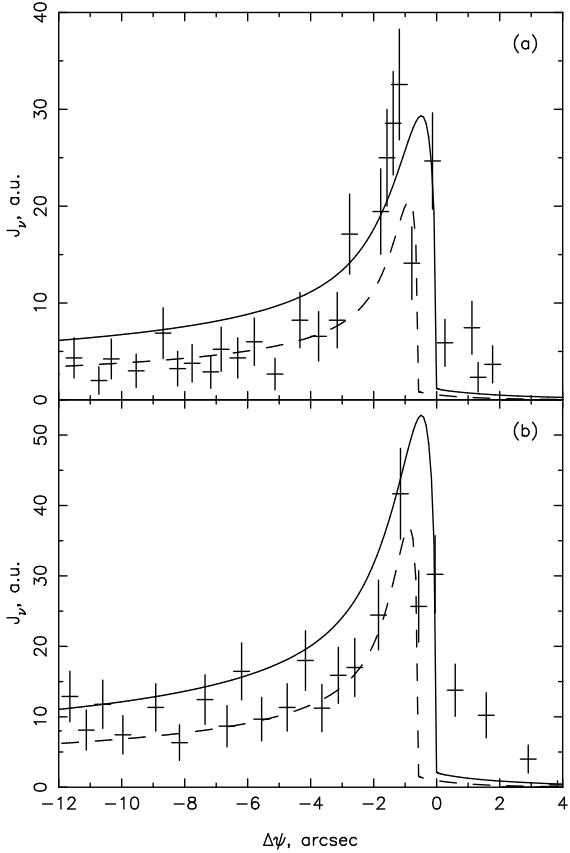}
\caption{Comparison of the low-field $B_\mathrm{d}=240$~$\mu$G
({\it solid curve}) and high-field $B_\mathrm{d}=360$~$\mu$G ({\it dashed
curve}) radial profiles of the full numerical solution in spherical
symmetry with the observational Chandra profiles (a) and (b) from
Fig.~\ref{f1}.  
}
\label{f3}
\end{figure}

The two extremes of the globally determined magnetic field from
Fig.~\ref{f2}
are then compared to the local Chandra data in Fig.~\ref{f3} by plotting the
full numerical solutions for the SNR morphology, as it results from the
combination of gas dynamics and time-dependent kinetic theory, together
with the data from Fig.~\ref{f1}. The agreement of global morphology and local
profiles is reasonable even though the differences between the two 
profiles show that the deviations from spherical symmetry are
significant. Such a comparison is of course incomplete since only two
radial profiles could be used. Further Chandra data are needed to
evaluate in detail the degree of agreement between theory and experiment.

\begin{figure}
\centering
\includegraphics[width=7.5cm]{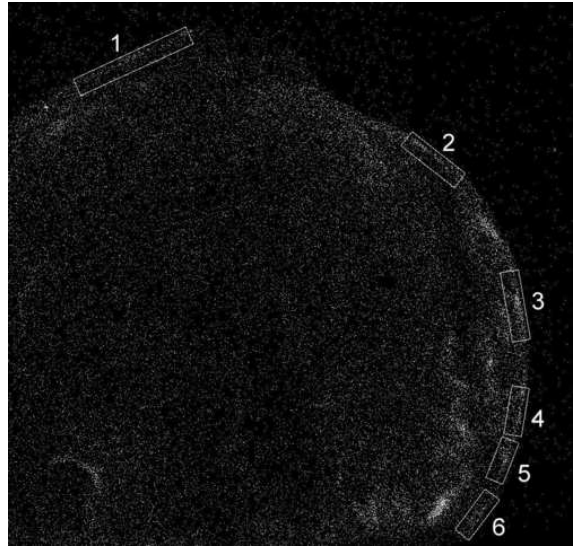}
\caption{Chandra ACIS spectroscopic array image of the northern and
northwestern part of \object{Tycho's SNR} in the 4 - 6 keV band. Shown are 6
selected boundary regions ({\it boxes}) that are assumed to contain 
portions of the
shock. Each region defines a radial profile from which a value of $B_\mathrm{d}$ is
derived. The sum of all 6 profiles defines an average profile which is
shown in Fig.~\ref{f5}.}
\label{f4}
\end{figure}

\begin{figure}
\centering
\includegraphics[width=7.5cm]{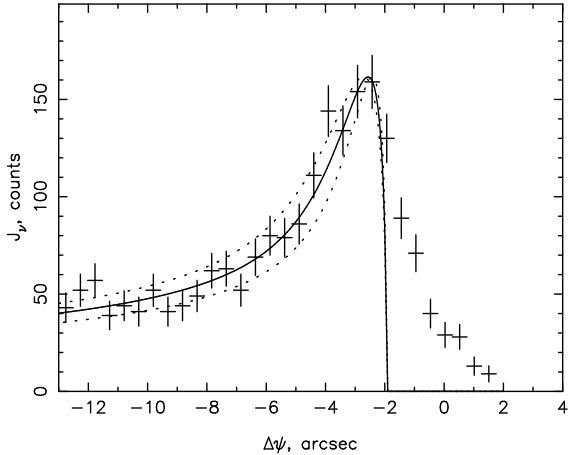}
\caption{Sum profile of the 6 radial X-ray profiles imaged in
Fig.~\ref{f4}. The
data are fitted to the projection of the exponential emissivity profile
described by the model of Eq. 3 in the downstream region ({\it solid
line}). The fit has a $\chi^2 = 9.73$, with $\chi^2/dof = 0.69$. The {\it
dotted lines} indicate the $1\sigma$ deviations.}
\label{f5}
\end{figure}

Finally, we have investigated Chandra X-ray images in the north and
northwest, obtained for X-ray energies $4 <\epsilon_{\nu} < 6$~keV with
the ACIS spectroscopic array from the Chandra archive, using the
morphology template. This corresponds to 6 profiles through the remnant's
outer boundary as indicated in the unsmoothed broadband Chandra ACIS image
in Fig.~\ref{f4}, showing a partial image of the SNR. In addition we show the
co-added data from the 6 profiles. The corresponding data are fitted to
the template. The individual results for $B_\mathrm{d}$ vary between 150$~\mu$G and
373$~\mu$G, with typical erros of (- 34 \% + 70 \%). Since the number of
events per bin is partly small, of the order of a few in the outer flanks
of the individual distributions, we have also coadded the profiles to
obtain a sum profile. It is shown in Fig.~\ref{f5}. Even though the individual
low-statistics profiles lead to a sizable scatter for $B_\mathrm{d}$, the average
profile determines a mean value of $B_\mathrm{d} = 273 (+ 49 - 37)$~ $\mu$G with a
very high fitting quality which agrees rather well with the optimized
field value from acceleration theory.

Note, that the actual shock position relative to the observed 
brightness profile is not known. Therefore it is used in the fitting 
procedure 
as a free parameter. As a result of the best fit the shock position 
corresponds to $\Delta \psi =0.3$ and 0.6~arcsec for the cases presented 
in Fig.~\ref{f1}a and \ref{f1}b respectively, and $\Delta \psi =-1.9$~arcsec for the case, 
presented in Fig.~\ref{f5}.
The overall situation is then as follows: the simple approximation
that determines an exponential width $L$ and then uses Eqs.(\ref{eq2})
and (\ref{eq3}) to
determine $l_2$, yields $B_\mathrm{d}$-values that agree well with each other {\it
and with the global spectral determination}. The same is true for the fit
of the profiles from the archive to the template given by Eq.(\ref{eq1}).  
Therefore we can also conclude that the coincidence of the postshock and
the global field values confirm that the field amplification during the
SNR evolution scales like $B_\mathrm{d}^2=const \times V_\mathrm{s}^2$.

\subsubsection{Gamma-ray prediction}

\begin{figure} 
\centering 
\includegraphics[width=7.5cm]{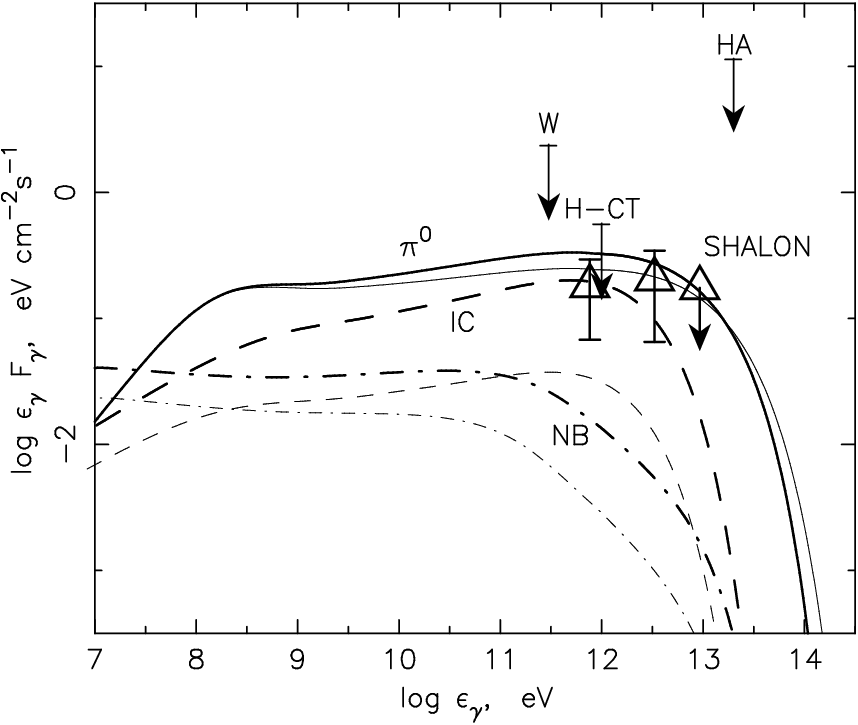}
\caption{Predicted \gr spectral energy distribution, as a function of \gr
energy: IC emission ({\it dashed lines}), Nonthermal Bremsstrahlung (NB,
{\it dash-dotted lines}), and $\pi^0$-decay ({\it solid lines}). The
observed $3\sigma$ \gr upper limits (W - Whipple (Buckley et al.
\cite{buc98}),
H-CT - HEGRA IACT system (Aharonian et al. \cite{aha01b})), and the 95\% confidence
HA - HEGRA AIROBICC upper limit (Prahl et al. \cite{pp97}) are shown together
with the detection/upper limit claimed by the Shalon group (Sinitsyna et
al. \cite{sin03}). {\it Thick lines} correspond to a downstream field of $B_\mathrm{d} =
240$~$\mu$G, whereas {\it thin lines} stem from assuming the maximum field
value $B_\mathrm{d} = 360$~$\mu$G.}
\label{f6}
\end{figure}

For completeness we shall also give the resulting prediction for the total
integral gamma-ray energy flux from \object{Tycho's SNR} for the above extreme
values of $B_\mathrm{d}$, equal to $240$~$\mu$G and $360$~$\mu$G, respectively, at
the same gas density (Fig.~\ref{f6}).  Compared to the earlier
prediction (V\"olk et al. \cite{vbkr02}), the hadronic $\pi^0$-decay flux changes
very little, as expected, whereas the Inverse Compton flux changes by a
factor of about 7 between $240$~$\mu$G and $360$~$\mu$G in effective
magnetic field strength, given the observed synchrotron flux. Thus a
detection of \object{Tycho's SNR} in TeV gamma rays close to the predicted hadronic
flux would again indicate a hadronic CR source - as the entire discussion
in terms of acceleration theory implies in any case. However, the hadronic
gamma-ray emission is quite sensitive to the ambient gas density which may
not have been fixed sufficiently well yet from radio observations (e.g.
Reynoso et al. \cite{rmg97}). We note however that the nominal value for the mean
ambient hydrogen density $N_\mathrm{H} = 0.5$~cm$^{-3}$, chosen in Fig.~\ref{f6}, 
is on the low rather than on the high side.

\subsection{\object{RCW~86}}

This SNR, more generically designated as G315.4-2.3, is a shell-type SNR
of large angular size $\sim 45$ arcmin, with a nonthermal radio spectrum,
at an apparent kinematic distance of $d=2.8$~kpc found from optical
observations (Rosado et al. \cite{ros96}). The distance might in fact be smaller,
and might be as low as $d\approx 1$~kpc (see e.g. Dickel et al.
\cite{dick};
Borkowski et al. \cite{bor01}, for more recent reviews). The designation
\object{RCW~86} is
an optical one and refers to a bright complex of optical emission
filaments in the southwest part of the more or less circular and complete
shell. This southwest part also exhibits the brightest radio and X-ray
emission. G315.4-2.3 has often been compared to \object{Tycho's SNR}. However, it
lacks the sharp outer rims that delineate much of \object{Tycho's SNR} silhouette (cf.
Dickel et al. \cite{dick}), in particular also in hard X-rays (Fig.~\ref{f4}).

The X-ray observations above 2 keV are well described by a dominant 
synchrotron continuum and a strong iron $K\alpha$~line. Using Chandra, Rho 
et al. (\cite{rho02}) in particular investigated spatial profiles of the hard 
($2<\epsilon_{\nu}<8$~keV) X-ray emission in the \object{RCW~86} region proper, one 
of which we consider appropriate to represent a strong shock since it has 
all the expected properties: a sharp rise in hard X-rays -- accompanied by 
a much smoother rise in the 1.4 GHz radio emission -- followed by a 
somewhat slower decline towards the same level of background. For this 
region in the west of \object{RCW~86} the 1.4 MHz emission profile has a more than 
ten times larger decay scale than the X-ray profile and may well be 
strongly influenced by adiabatic expansion effects, masking the 
comparatively weak radiative losses at those electron energies. We shall 
concentrate on this single profile which in any case yields the largest 
value of the effective field available for this source. The data for this 
profile appear to also have been used by Bamba et al. (\cite{bam03b}) and Bamba 
(\cite{ba04}) in their analysis. Their exponential angular width leads to 
$L=0.29$~pc for the sharpest profile, that is very close to our estimate 
$L=0.26$~pc from the Rho et al. (\cite{rho02}) profile. However, even this profile 
is in the interior of \object{RCW~86}, not at the rim. It is possible that this 
part of the shock surface has a curvature due to an inhomogenuity of 
surrounding medium which is quite different from that of an ideal 
spherical surface. We therefore have to consider the likelihood of 
nonspherical geometry and thus of incomplete de-projection with a ratio 
$L/l_2 <7$, as in the case of \object{RX~J1713.7-3946} below.  Using 
nevertheless Eq.(\ref{eq2}) for the time being, we obtain $B_\mathrm{d} \approx 108$~$\mu$G 
for our estimate from the Rho et al. data and $B_\mathrm{d} \approx 100$~$\mu$G for 
the Bamba et al. width. The shock has been assumed to move with a velocity 
of $V_\mathrm{s}=800$~km/sec into a low-density medium with hydrogen
density $N_\mathrm{H} = 
0.3$~cm$^{-3}$ at a kinematic distance of $d=3$~kpc, cf. Borkowski et al.  
(\cite{bor01}).

\begin{figure} 
\centering 
\includegraphics[width=7.5cm]{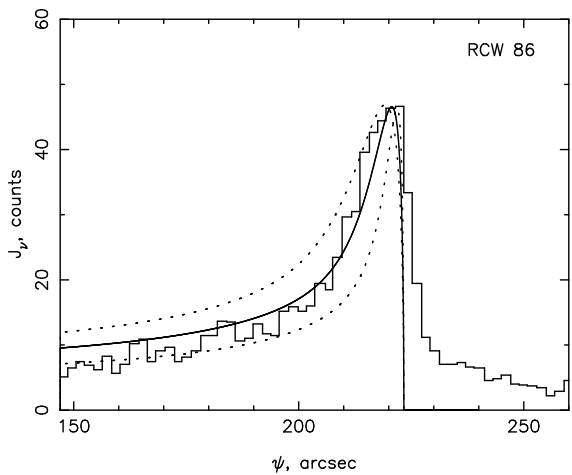}
\caption{Fit of the data ({\it histogram})
corresponding to the Rho et al. (\cite{rho02}) profile
from Chandra in the 2 - 8 keV X-ray band for \object{RCW~86} to the template cf.  
Eq.(\ref{eq1}) ({\it solid line}). The $1 \sigma$ deviations are given by the {\it
dotted curves.}}
\label{f7}  
\end{figure}

If we apply the model fit cf. Eq.(\ref{eq1}) to the top profile of Fig.~8 of Rho 
et al. (\cite{rho02}) we obtain $B_\mathrm{d} \approx 99 (+ 46 - 26)$~$\mu$G with a 
$\chi^2$/dof = 0.35, using 25 dof and assuming spherical geometry  
as our preferred value for full de-projection (see Fig.~\ref{f7}).

If we were to assume a distance of $d=1$~kpc to \object{RCW~86} instead, the linear 
scales would decrease by a factor of about 3, increasing the effective 
field by a factor of about 2 to $B_\mathrm{d} = 231$~$\mu$G. Whereas a value of 
$B_\mathrm{d}=100$~$\mu$G corresponds to a comparatively modest field amplification 
of about a factor of 5, we must therefore count on a very strong 
amplification in the case that the source is nearby. This would be no 
surprise, assuming that for the close distance the source might be as 
young as 2000~yr, instead of $\sim 10^4$~yr for the large distance 
(Borkowski et al. \cite{bor01}), resulting in a high shock velocity and 
correspondingly expected $P_\mathrm{c} \sim \rho V_\mathrm{s}^2$.

In reality however we must count on somewhat larger scales $l_2$ of the 
emissivity for the observed brightness scale $L$. A ratio $l_2/L = 1/4$ or 
even $l_2/L = 1/2$ would reduce the above field amplification by a factor 
0.69 or even 0.43, respectively. These appear realistic correction 
factors, given the fact that the shock transition lies in the outer part 
of the shell.

\subsection{\object{Kepler's SNR}}

\object{Kepler's SNR} is 400 yr old and has an angular radius of $\approx
100$~arcsec at a distance $d\approx 4.8 \pm 1.4$~kpc (Reynoso \& Goss
\cite{rg99}). This implies a linear radius $R_\mathrm{s}\approx 2.3 \pm 0.7$~pc, somewhat
smaller than \object{Tycho's SNR}. Like the optical and the radio continuum
emission the X-ray brightness distribution from \object{Kepler's SNR} is rather
asymmetric with stronger emission in the north and northwest (e.g. DeLaney
et al. \cite{deL};  Cassam-Chena$\ddot{\imath}$ et al. \cite{cas04a}). The most obvious
nonthermal hard X-ray continuum emission in the 4 - 6 keV range, as
observed with XMM-Newton, is found in a southeastern region, rather at the
opposite side of the remnant (Cassam-Chena$\ddot{\imath}$ et al. \cite{cas04a}),
where also thin filaments have been observed with Chandra. The spectra of
these filaments have been found to be hard without line-structures by
Bamba et al. (\cite{bam03b}) and by Bamba (\cite{ba04}), from Chandra archival 
data, with the conclusion that the filaments are the result of 
synchrotron emission. One profile has again the characteristics of a 
shock transition, for which they show a scale length of about 
$L=0.11$~pc in the 4.0-10.0 keV band.

Applying our de-projection model to this source despite its asymmetry, 
we obtain $l_2 \approx 1.6 \times 10^{-2}$~pc.  
Cassam-Chena$\ddot{\imath}$ et al.  (\cite{cas04a}) identify this southeastern 
part of the SNR as a very low density medium with $N_\mathrm{H} \leq 
0.15$~cm$^{-3}$ and a shock velocity $V_\mathrm{s} \approx 4550$~ km/s. Using 
this velocity in Eq.(\ref{eq3}) for $\delta$, and assuming an overall shock 
compression ratio of $\sigma=6$, we obtain from Eq.(\ref{eq2}) the 
substantial field $B_\mathrm{d} \approx 215$~$\mu$G, with an estimated 
error of less than 20\%. Thus we conclude that there is 
substantial field amplification in \object{Kepler's SNR}.

\subsection{SNR \object{RX~J1713.7-3946}}

This very large $\approx 1$~degree diameter shell-type SNR in the Galactic
plane was discovered in the ROSAT all-sky survey (Pfeffermann \&
Aschenbach \cite{pfe96}). The hard X-ray emission detected with ASCA turned out to
be purely nonthermal and did not reveal any thermal emission (Koyama et
al. \cite{koya97}; Slane et al. \cite{sla99}). Recently a weak thermal component may have
been detected in the interior of the remnant using large field-of-view
X-ray instruments, including the RXTE Proportional Counter Array (PCA)
(Pannuti et al. \cite{pann03}).  The CANGAROO experiment reported a detection in
TeV \grs at a level of 70\% of the Crab Nebula (Muraishi et al.
\cite{mur00}; Enomoto et al. \cite{enomoto}).

The H.E.S.S. experiment has confirmed (Aharonian et al. \cite{aha04}) the CANGAROO
detection. The gamma-ray and X-ray images have similar shapes. There
appears to be no doubt about the reality of this source in TeV gamma-rays.

The SNR is complex. New CO-observations by Fukui et al. (\cite{fukui}) suggest the
source to be rather close, at a distance of about $d=1$~kpc. A short
distance of $d=1.3 \pm 0.4$~kpc has recently also been inferred by
Cassam-Chena$\ddot{\imath}$ (\cite{cas04b}) from an a detailed investigation of
the X-ray absorbing material in the remnant together with the
CO-observations.

Hard X-ray ($1<\epsilon_{\nu}<5$~keV) observations with Chandra by 
Uchiyama et al. (\cite{uch03})  and Lazendic et al. (\cite{laze}) revealed filaments in 
the remnant interior. They appear embedded in a diffuse plateau emission 
(Cassam-Chena$\ddot{\imath}$ \cite{cas04b}). According to Uchiyama et al. the most 
prominent and largest filament has a very small width of about 20 arcsec. 
For a distance of $d=1$~kpc this translates into a spatial scale of 
$L=0.1$~pc. At a different portion of this filament Lazendic et al. (\cite{laze}) 
inferred a width of about 40 arcsec, i.e. twice as large. We note however 
that the width drived from the observed profiles depends appreciably on 
the assumed background level. For a lower background level than that 
assumed by Uchiyama et al. (\cite{uch03})  and Lazendic et al. (\cite{laze}) the 
experimental width may be considerably higher, with a corresponding 
magnetic field value that is considerably lower than estimated below.

It is likely but not certain that the filament measured corresponds to a
shock. A tangential discontinuity due to an internal structure of the
circumstellar medium before the explosion or a discontinuity separating
shocked ejecta from the shocked circumstellar medium cannot be excluded.
To this physical possibility one must add the probability of shear-type
three-dimensional convective flows, distinct from spherical expansion. In
particular, the projection of such a structure along the line of sight
might be far different in scale from that for a spherical structure with a
radius comparable to the dimension of the overall object. It is indeed
quite possible that the Chandra profile cuts have not crossed the outer
shock, and therefore do not necessarily delineate the main field
amplification region caused by the accelerated particles.

Nevertheless, we can attempt some estimates for different possible shock 
situations regarding the observed filament: (i) a shock transition of 
scale $L=0.1$~pc, assuming a compression ratio $\sigma = 5$ and a speed of 
$V_\mathrm{s}=4000$~km/s, and a low upstream density $N_\mathrm{H} = 2\times 
10^{-2}$~cm$^{-3}$ (Cassam-Chena$\ddot{\imath}$ \cite{cas04b}) for an X-ray energy 
$\epsilon_{\nu} = 1 $~keV gives a maximum $B_\mathrm{d} \approx 271$~$\mu$G; it 
reduces to $B_\mathrm{d} \approx 74$~$\mu$G without any de-projection, $l_2 = L$, 
and to $B_\mathrm{d} \approx 118$~$\mu$G for $l_2 = L/2$; (ii) for a shock in a 
denser medium $N_\mathrm{H} \approx 1.3$~cm$^{-3}$, with velocity $V_\mathrm{s} = 
1000$~km/s, the maximum $B_\mathrm{d}= 213$~$\mu$G; it reduces to $B_\mathrm{d} \approx 
58$~$\mu$G for $l_2 = L$, and to $B_\mathrm{d} = 94$~$\mu$G for $l_2 = L/2$. For 
the twice thicker Lazendic et al. filament the foregoing field values 
would reduce by a factor of about 0.6. This range of field strengths might 
provide an idea what to expect from interior structures inside this SNR. 
New high-resolution measurements appear necessary for the global picture.

\subsection{Cas~A}

Extending Bamba et al. (\cite{bam03b}), Bamba (\cite{ba04}) gives two Chandra profiles 
through the southeastern rim at X-ray energies $5 <\epsilon_{\nu}<10$~keV 
whose average width is $L\approx 0.028$~pc and which we convert to $l_2 
\approx 4 \times 10^{-3}$~pc. Using the hydrodynamic solution of Berezhko 
et al. (\cite{bpv03}) for the remnant dynamics, with a shock compression ratio 
$\sigma \approx 6$ and a shock speed $V_\mathrm{s} =2400$~km/s, we obtain $B_\mathrm{d} 
\approx 485$~$\mu$G, taking $\epsilon_{\nu}= 5 $~keV.  If we were to take 
$V_\mathrm{s} =3000$~km/s we would obtain $B_\mathrm{d} \approx 496$~$\mu$G.

Vink \& Laming (\cite{vl03}) have analyzed a profile with $L \approx 0.025$~pc 
through the northeastern rim at $4<\epsilon_{\nu}<6$~keV, and assumed a 
shock velocity of $V_\mathrm{s}=5000$~km/s. For $V_\mathrm{s} =3000$~km/s and using this 
scale $L$, we calculate $B_\mathrm{d} \approx 550$~$\mu$G (Berezhko \& V\"olk 
\cite{bv04a}).

These values from the two different profiles agree quite well with each 
other. They differ by less than 15 \% from the spectrally 
determined value $B_\mathrm{d} \approx 480$~$\mu$G from Berezhko et al.
(\cite{bpv03}).

\subsection{SN 1006}

For SN 1006 Bamba et al. (\cite{bam03a}, \cite{bam04}) for $\epsilon_{\nu}>2$~keV give 
exponential widths for the sharpest rim profile $L=19$~arcsec, which for 
the distance $d=2.2$~kpc translates into $L=0.2$~pc.  Using a shock 
velocity $V_\mathrm{s} = 3200$~km/s and a shock compression ratio $\sigma = 6.3$, 
Eq.(\ref{eq2}) yields a downstream field value $B_\mathrm{d} \approx 143$~$\mu$G.

According to Bamba et al. (\cite{bam03a}, \cite{bam04}) and Bamba
(\cite{ba04})  there is an even 
sharper profile with $L= 0.126$~pc, which is however not a rim profile. 
With $l_2=L/7$ this leads to $B_\mathrm{d} \approx 195 $~$\mu$G.

These numbers can be compared to the average field value 120~$\mu$G 
predicted from the overall synchrotron spectrum by Berezhko et al.  
(\cite{bkv02}), and depending very weakly on the ambient gas density. In fact, 
reconsidering the spectral determination of $B_\mathrm{d}$, we can achieve 
a preferred field value of 160~$\mu$G (Ksenofontov et al. \cite{kse04}).

Taking into account the data uncertainty, which is about 25\%, we 
can conclude that the above numbers derived from the Bamba et 
al. (\cite{bam03a}, \cite{bam04}) data are consistent with the values derived from the 
synchrotron spectrum. Together they confirm the overall field 
amplification picture for SN 1006.  On the other hand, there may 
also be a physical reason for the differences in the values of $B_\mathrm{d}$ 
across the shock surface. This might occur if in the polar caps 
independent magnetic flux tubes exist, between which the field 
strength and probably also the generating CR pressure $P_\mathrm{c}$ can 
fluctuate considerably.

\section{Discussion}

All four newly calculated sources show amplified fields. In addition
\object{Tycho's SNR} furnishes a third case where the field strength $B_\mathrm{d}$
determined from small-scale synchrotron features agrees with that
predicted by comparing the global synchrotron spectrum with nonlinear
acceleration theory.

Although the field amplifications may be relatively small for
\object{RCW~86} and
\object{RX~J1713.7-3946} and are difficult to pin down, all six young Galactic
SNRs with known filamentary scales exhibit field amplification. There
appears to be no exception to this ``rule'' up to now.

As an important prerequisite the characteristic nonlinear emission
features require a {\it dominant nuclear component} of accelerated
particles. At the moment this conclusion is in part still a theoretical
one: although the \gr flux at 1 TeV from Cas~A, detected by the HEGRA
experiment (Aharonian et al. \cite{aha01a}), agrees with the theoretically
calculated hadronic flux there is as yet no confirmation by another
experiment. SN 1006, reported by the CANGAROO experiment (Tanimori et al.
\cite{tan98}, \cite{tan01}) could not be detected by the H.E.S.S. experiment as a TeV
source until now (Masterson et al. \cite{mas03}; Rowell et al.
\cite{row04}). An inverse
Compton \gr emission scenario in a low magnetic field of order $B_\mathrm{d}\approx
10$~$\mu$G (e.g. Pohl \cite{poh96}, Mastichiadis \& de Jager \cite{mas96}; Tanimori et al.
\cite{tan98}, \cite{tan01}) is clearly excluded by the H.E.S.S. upper limits.  The
H.E.S.S. upper limit is almost one order of magnitude lower than the
published CANGAROO flux. Given the range of ambient densities $0.05 \leq
N_\mathrm{H} /(1~\mbox{cm}^{-3}) \leq 0.3$ it suggests a low density ISM
with $N_\mathrm{H}\le
0.1$~cm$^{-3}$ (Ksenofontov et al. \cite{kse04}) instead of the
high value $N_\mathrm{H}= 0.3$~cm$^{-3}$, which happened to fit the CANGAROO data
(Berezhko et al. \cite{bkv02}). Finally, the predicted hadronic gamma-ray flux for
\object{Tycho's SNR} (V\"olk et al. \cite{vbkr02}) is still below the upper limit measured
by HEGRA (Aharonian et al.  \cite{aha01b}). Only
\object{RX~J1713.7-3946}
has been detected by two independent gamma-ray experiments.

In all cases the gamma-ray emission depends strongly on "external"
astronomical parameters, like the ambient gas density, and therefore
requires extensive multi-wavelength investigations with significant
inherent uncertainties.

Regarding the magnetic field strengths derived we can attempt to learn
something from these results by empirically correlating the magnetic field
pressure $B_\mathrm{d}^2/(8\pi)$ with the ram pressure $\rho_0 V_\mathrm{s}^2$ of the
upstream gas in the shock frame of reference. Assuming that $B_\mathrm{d}^2/(8\pi)$
reaches some fraction of the CR pressure $P_\mathrm{c}$ at the shock, and assuming
very efficient acceleration so that $P_\mathrm{c}= \epsilon \rho_0 V_\mathrm{s}^2$, with
$\epsilon = O(1)$, we expect a constant ratio $B_\mathrm{d}^2/(8\pi \rho_0 V_\mathrm{s}^2)$
as a function of shock velocity $V_\mathrm{s}$ for an individual object. This
appears to us to be the physically most natural result, given the overall
dynamics in the shock (see also the discussion in V\"olk et al. \cite{vbkr02}).
Berezhko \& V\"olk (\cite{bv04b}) have used this assumption to calculate the
proton and electron spectra for a generic SNR with field amplification.
The result is that the proton component reaches the so-called knee in the
Galactic CR spectrum with a maximum momentum $p_\mathrm{max} \sim 10^6
m_\mathrm{p}$c.

In numbers the ratio of magnetic and kinetic energy densities is given 
by:
\begin{equation} 
\frac{B_\mathrm{d}^2/(8\pi)}{\rho_0 V_\mathrm{s}^2} \approx
1.7 \times 10^{-2}\left\{
\frac{[B_\mathrm{d}/(100~\mu\mathrm{G})]^2}
{[N_\mathrm{H}/(1~\mathrm{cm}^{-3})][V_\mathrm{s}/(10^3~\mathrm{km/s})]^2}
\right\},
\end{equation} 
where $B_\mathrm{d}$ may itself implicitly depend on $V_\mathrm{s}$ if it is determined
from Eqs.(\ref{eq2}) and (\ref{eq3}).

On the other hand, Bell (\cite{bell04}) has argued for a proportionality 
$B_\mathrm{d}^2/(8\pi) \propto (V_\mathrm{s}/c) \rho_0 V_\mathrm{s}^2$ from his quasi-MHD simulations 
of the nonresonant right-hand polarized low frequency MHD mode. This 
implies a linear increase of $B_\mathrm{d}^2/(8\pi \rho_0 V_\mathrm{s}^2)$ with $V_\mathrm{s}$. Bell 
has suggested that acceleration even considerably beyond the knee may be 
possible in this way in very young SNRs expanding into dense circumstellar 
material. This extends earlier arguments put forward by Bell \& Lucek 
(\cite{bluc01}) who estimated the maximum achievable energy $c p_\mathrm{max}$. However, 
the results of Berezhko \& V\"olk (\cite{bv04b}) indicate that the energy content 
of the fastest part of the supernova ejecta, which produces CRs with 
energies above the knee energy $3\times 10^{15}$~eV, is so small that the 
resultant CR spectrum produced in SNRs is very steep $N\propto 
\epsilon^{-5}$. Therefore these highest energy CRs play no signigifant 
role in the formation of the Galactic CR spectrum above the knee. The 
reason is the shortness of the time period during the initial SNR phase 
over which a very high energy flux density $\rho {V_\mathrm{s}^3}/2$ into the shock 
is maintained.

\begin{figure} 
\centering 
\includegraphics[width=7.5cm]{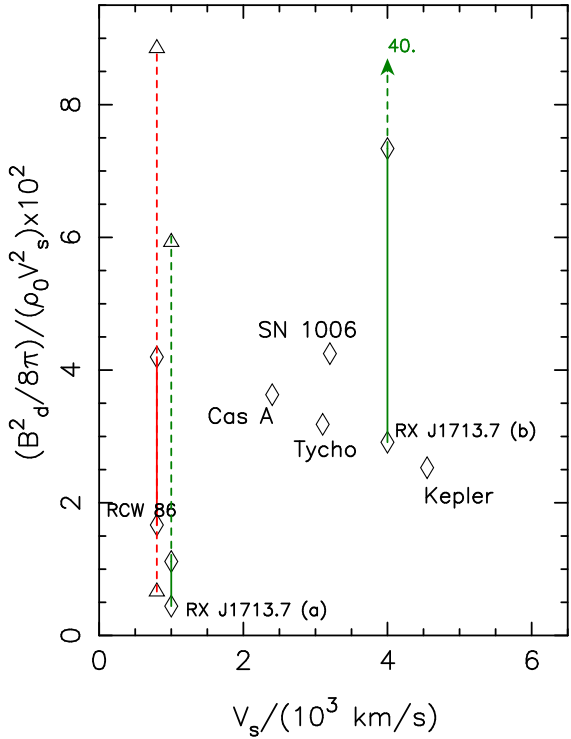}
\caption{Ratio of downstream magnetic field pressure $B_\mathrm{d}^2/(8\pi)$ to
preshock gas ram pressure $\rho_0 V_\mathrm{s}^2$ (in percent) vs. shock velocity
$V_\mathrm{s}$ (in units of $10^3$~km/s) for the six young SNRs that show field
amplification. For the objects \object{RCW~86} and
\object{RX~J1713.7-3946} a range of
normalized magnetic field pressures is given.
(For \object{RCW~86} the four points
correspond to $L/l_2 = 1$, 2, 4 and 7, respectively, and
the {\it solid vertical line segment} corresponds to the assumed range of
uncertainty $7/4 < 7 l_2/L < 7/2$. For \object{RX~J1713.7-3946} the three
points correspond to $L/l_2=1$, 2 and 7, respectively, and
the {\it solid vertical line segment} corresponds to the uncertainty range
$7/2 < 7 l_2/L < 7$) For this remnant two alternative shock states
are depicted: the low velocity case (a) corresponds rather to the cloudy
medium in the north and west of the remnant. The high velocity case (b) is
more representative for the very low density regions in the southern
projection of the remnant, the fully projected case $L/l_2 = 7$
corresponding to a value of 40, out of scale in the figure.}
\label{f8}
\end{figure}

Obviously we can not follow an individual source as it evolves in time.  
The only possibility is to make the strong assumption that different
objects at different epochs of their expansion history behave like a
single source at such points in its evolution. I.e. we can substitute an 
ensemble average for a time average. Given that we consider here even 
different types of SN explosions (Type Ia for \object{SN~1006},
\object{Tycho}, and possibly 
\object{Kepler}, probably type II for \object{RX~J1713.7-3946} and
\object{RCW~86}, and probably type 
Ib for \object{Cas A}), conclusions must be drawn with this fact in mind.

Fig.~\ref{f8} shows a plot of $B_\mathrm{d}^2/(8\pi \rho_0 V_\mathrm{s}^2)$ vs. $V_\mathrm{s}$ for the six 
sources discussed in the previous section. For \object{Tycho's SNR},
\object{SN~1006}, and \object{Cas A} the available globally determined field values were 
used. For the sources with rim profiles (\object{Tycho's SNR},
\object{SN~1006}, 
\object{Kepler's SNR}, and \object{Cas A}) we have assumed the full de-projection $L = 7 
l_2$, and $N_\mathrm{H}=0.1$~cm$^{-3}$ and $N_\mathrm{H}=2$~cm$^{-3}$
for \object{SN~1006} and 
\object{Cas~A}, respectively. 

For \object{RCW~86} we have assumed a range of uncertainty $7/4 < 7 l_2/L < 7/2$ 
(the four points in Fig.~\ref{f8} correspond to $L/l_2 = 1$, 2, 4 and 7, 
respectively), and for \object{RX~J1713.7-3946} we have taken the even more 
pessimistic range $7/2 < 7 l_2/L < 7$ (the three points in Fig.~\ref{f8} 
correspond to $L/l_2=1$, 2 and 7). These values correspond to the 
discussion of these two sources in the previous section.

For \object{RX~J1713.7-3946} we have in addition plotted the two alternatives 
corresponding to $V_\mathrm{s} = 1000$~km/s and $N_\mathrm{H} \approx 1.3$~cm$^{-3}$ on the 
one hand, and $V_\mathrm{s} = 4000$ km/s and $N_\mathrm{H} \approx 2 \times 
10^{-2}$~cm$^{-3}$ on the other.

The present uncertainties in the data, which are not more accurate
than  25\% and cover only a rather limited range of the shock speed 
$V_\mathrm{s}$, do not allow a distinction between a constant ratio
$B_\mathrm{d}^2/(8\pi  \rho_0 V_\mathrm{s}^2)$ and one with a linear
growth of $B_\mathrm{d}^2/(8\pi \rho_0  V_\mathrm{s}^2)$ with
$V_\mathrm{s}$. However, the data are highly likely to exclude a
decrease  of $B_\mathrm{d}^2/(8\pi \rho_0 V_\mathrm{s}^2)$ with
increasing $V_\mathrm{s}$. Obviously this  means that with growing age,
that is decreasing shock velocity, the  field amplification in SNRs
decreases about linearly with the shock  velocity. Fig.~\ref{f8}
indicates that in all young SNRs with  large enough shock
speed $V_\mathrm{s}>10^3$~km/s the effective (amplified)  magnetic
field energy density is near {\bf $B_\mathrm{d}^2/(8\pi)=3.5\times 
10^{-2}\rho_0V_\mathrm{s}^2$}, the value for \object{Kepler's SNR} probably having the
largest  uncertainty.  It is important to note that in all four of the
best  defined cases (\object{SN~1006}, \object{Cas~A}, \object{Tycho's
SNR} and \object{Kepler's SNR}), which are 
characterized by rather close values of the shock speed, the
normalized  magnetic field pressures, extracted from the data, are also
close to  each other. The large uncertainties in the determined
magnetic field  values for the cases of \object{RCW~86} and
\object{RX~J1713.7-3946} are
due to much more poorly  known values of their relevant parameters,
such as age, distance, shock  speed and ISM gas density.

We thus tentatively conclude that field amplification is confined to young
remnants. This also implies that the escape of the highest energy nuclear
particles, accelerated at an earlier epoch with high effective field,
becomes progressively important as the remnant age increases. This implies
a lower and lower cutoff of the resulting gamma-ray spectrum from a SNR
with age. As a result, the highest energy nuclear particles are to be
found in SNRs that have just reached the Sedov phase. Older remnants are
increasingly unable to confine them.

The amplified magnetic field also leads to a depression of the Inverse
Compton and nonthermal Bremsstrahlung gamma-ray emission relative to the
hadronic emission for young remnants, given the synchrotron emission.
Nevertheless the hadronic gamma-ray emission depends strongly on
$N_\mathrm{H}$,
essentially $\propto N_\mathrm{H}^2$ (V\"olk \cite{voel03}). Thus a strong source of nuclear
CRs can still be a gamma-ray source with a comparable or even preponderant IC
emission. Ultimately even SN 1006 might turn out as an example for that
case, since its $N_\mathrm{H}$ must be quite small .

\section{Conclusions}

From the fact that magnetic field amplification occurs in all the young
SNRs for which relevant data exist, and given the strong theoretical
connection between magnetic field amplification and efficient acceleration
of nuclear CRs, we tentatively conclude that the Galactic SNRs are the
source population of the Galactic CRs.

Quantitatively, with \object{Cas~A}, \object{SN~1006}, and now
\object{Tycho's SNR} from this 
paper, there exist three examples where nonlinear acceleration theory 
and X-ray observations of filamentary structures give the same 
morphology and the same magnetic field amplification effect: the 
effective interior magnetic field energy density is near {\bf 
$B_\mathrm{d}^2/(8\pi)=3.5\times 10^{-2}\rho_0V_\mathrm{s}^2$}. The estimates made for 
the cases of \object{Kepler's SNR}, \object{RCW~86} and
\object{RX~J1713.7-3946} confirm such a 
conclusion although less definitely so due to their less well-defined 
parameter values.

The data suggest that field amplification and thus the acceleration of 
nuclear particles to the highest energies that correspond to the knee in 
the observed CR spectrum (Berezhko \& V\"olk \cite{bv04b}), or even possibly 
beyond (Bell \& Lucek \cite{bluc01}; Bell \cite{bell04}), is a transitory effect in CR 
sources, limited to the early evolutionary phase of SNRs. This agrees with 
theoretical expectations. From acceleration theory the later phases of SNR 
evolution, which are characterized by decreasing shock speeds, should then 
maintain the lower energy particle population produced early on, in a 
balance between adiabatic expansion and continuing acceleration.

The highest-energy gamma-rays emitted from the CR sources should be
observable only during the early phases of evolution, roughly at the end
of the SNR sweep-up phase, before the highest energy particles have already
left the source.

\begin{acknowledgements}
We thank A.\ R.\ Bell and G.\ P\"uhlhofer for critical discussions.
This work has been supported in part by the Russian Foundation for Basic
Research (grant 03-02-16524). EGB acknowledges the hospitality of the
Max-Planck-Institut f\"ur Kernphysik, where this work was carried out.
\end{acknowledgements}

\end{document}